# A comparative study of two molecular mechanics models based on harmonic potentials

Junhua Zhao[1, 2*], Lifeng Wang[3], Jin-Wu Jiang[2], Zhengzhong Wang[1], Wanlin Guo[3*], Timon Rabczuk[2*]

[1]*Department of Materials and Structural Engineering, Northwest A&F University, 712100 Yangling, Shaanxi, PR China*

[2]*Institute of Structural Mechanics, Bauhaus-University Weimar, 99423 Weimar, Germany*

[3]*State Key Laboratory of Mechanics and Control of Mechanical Structures, Nanjing University of Aeronautics and Astronautics, 210016 Nanjing, PR China*

**Abstract**

We show that the two molecular mechanics models, the stick-spiral and the beam models, predict considerably different mechanical properties of materials based on energy equivalence. The difference between the two models is independent of the materials since all parameters of the beam model are obtained from the harmonic potentials. We demonstrate this difference for finite width graphene nanoribbons and a single polyethylene chain comparing results of the molecular dynamics (MD) simulations with harmonic potentials and the finite element method with the beam model. We also find that the difference strongly depends on the loading modes, chirality and width of the graphene nanoribbons, and it increases with decreasing width of the nanoribbons under pure bending condition. The maximum difference of the predicted mechanical properties using the two models can exceed 300% in different loading modes. Comparing the two models with the MD results of AIREBO potential, we find that the stick-spiral model overestimates and the beam model underestimates the mechanical properties in narrow armchair graphene nanoribbons under pure bending condition.

---

* Corresponding author. Tel: +49 3643 584506; +86 25 8489 1896; +49 3643 584511
  *Email address*: (JZ) junhua.zhao@uni-weimar.de; (WG) wlguo@nuaa.edu.cn; (TR) timon.rabczuk@uni-weimar.de

**1. Introduction**

Harmonic potentials have been extentively used to investigate the mechanical and physical properties of various materials in molecular mechanics models, such as carbon nanotubes (CNTs), boron nitride nanotubes (BNTs), graphene sheets and polymers (Chopra et al., 1995; Chang and Gao, 2003; Li and Chou, 2003; Zhao et al., 2010; Zhao et al., 2011; Jiang and Guo, 2011). Atomistic-based methods such as classical MD (Iijima et al., 1996; Yakobson et al., 1996), tight-binding MD (Hernandez et al., 1998; Zhao et al., 2009a), and density functional theory (Sanchez-Portal et al., 1999; Zhang et al., 2007; Zhang and Guo, 2008) have been used to study the mechanical properties of CNTs, BNTs and nanoribbons. However, compared with bottom-up approaches, top-down approaches may substantially reduce the computational costs and are thus frequently used in related investigations. Recently, the molecular dynamics (MD) simulation with harmonic potentials coupling finite element (FE) method have been more and more applied to multiscale modeling in order to characterize the mechanical behavior of the different materials from nanoscale to microscale/macroscale (Badia et al., 2007; Di Matteo et al., 2007; Bian and Wang, 2011), so the predictive ability based on the harmonic potentials has special importance.

Some typical continuum models based on the harmonic potentials have been developed and broadly used to clarify the elastic properties of the graphene sheets, CNTs and BNTs (Hernandez et al., 1998; Vaccarini et al., 2000; Ru, 2001; Chang and Gao, 2003; Li and Chou, 2003). Three kinds of models are usually employed: 1). Shell models have been used to capture the buckling characterizes of CNTs (Yakobson et al., 1996; Ru, 2000; Ru, 2001; He et al., 2005; Wang et al., 2007; Wu et al., 2008). The applicability and limitations of shell models have been extensively discussed (Wang, 2004; Peng et al., 2008; Zhang et al., 2009). Chang (2010) developed an anisotropic shell model to investigate mechanical behavior of single-walled CNTs, in which the model can be used to effectively describe the chirality effect on mechanical properties. 2). The beam model was developed by Li and Chou (2003, 2004). They assume that the beam elements have circular cross sections and are always subjected to pure tension, pure bending, and pure torsion. The theory was further improved (Tserpes and Papanikos, 2005; Xia et al., 2005; To, 2006; Kasti, 2007; Jiang et al., 2009) and extended to calculate the five independent size- and chirality-dependent elastic moduli of single-walled CNTs using equivalent beam elements with rectangular section (Li and Guo 2008). 3). The "stick-spiral" model (SSM) was developed by Chang and Gao (2003). An improved model by Jiang and Guo (2011) was used to investigate the elastic properties of single-walled boron nitride nanotubes. By extending the two analytical methods to crystalline polymers (Zhao et al., 2010), we presented the SSM to investigate the size-dependent elastic properties of crystalline polyethylene (PE) (Zhao et al., 2011). Based on the united-atom MD simulations, we further verified the effectivity of the SSM in the crystalline polymers directly (Capaldi et al., 2004; Zhao et al., 2011). In this work, we utilized a united atom approximation in which the methyl groups ($CH_2$) are represented by a single "atom" or unit, and the effect of the hydrogen atoms on the polymer's configuration is accounted for in the potentials (Waheed, 2005; Shepherd, 2006). Subsequently, we extended the beam-spring model to obtain the elastic properties of crystalline PE (Zhao et al., 2012).

Kasti (2007) found that the beam bending stiffness (BBS) ($EI/b$, where $E$ is the Young's modulus, $b$ is the beam length and $I$ is the moment of inertia of the beam (see Section 2)) is equal to the bond bending stiffness ($K_\theta$, which is the bond angle bending force constant (see Section 2)) in zigzag CNTs, while the BBS beam bending stiffness is only half of the bond bending stiffness $K_\theta$ in graphene nanoribbons. This discovery was verified in the zigzag CNT and graphene sheet based on energy equivalence.

Although the SSM and the beam models have been effectively used to describe the elastic properties of CNTs, BNTs and graphene sheets (Li and Chou, 2003, 2004; Kasti, 2007; Li and Guo 2008; Jiang and Guo, 2011), the difference of their prediction ability has never been systematically studied.

In this paper, we study the mechanical properties of the finite width graphene nanoribbons under different loading conditions using the two models. First, we consider the SSM under difference loading conditions. Then, the BBS of the graphene nanoribbons is derived from the energy equivalence between the two models. We show that the BBS strongly depends on the loading modes and the chirality in the finite width graphene nanoribbons. The closed-form expressions of the bending stiffness are derived under uniform tension, pure shear, pure bending, loading force, coupling force and bending conditions. Moreover, the BBS of the beam model under different loading conditions is systematically studied in the graphene nanoribbons using the MD simulation with present harmonic potentials (Chang and Gao, 2003) and the FE simulation. Finally, the results of the MD simulation with harmonic potentials and the FE method are compared with those of the MD results with AIREBO potential.

The paper is organized as follows: Section 2 describes the SSM and the beam model in armchair and zigzag graphene nanoribbons for different loading conditions. In Section 3, both models are validated by comparison to MD simulations and FE results. Moreover, a single PE chain under different loading conditions is investigated. The comparison of the two models with MD simulations using the AIREBO potential is discussed in Section 4. The paper is concluded in Section 5.

**2. The stick-spiral and beam models in graphene nanoribbons**

In the framework of molecular mechanics, the total energy, $U$, of graphene at small strains can be expressed as a sum of energies associated with the varying bond length, $U_b$, and bond angle, $U_\theta$, i.e., (Chang and Gao, 2003)

$$U = U_b + U_\theta = \frac{1}{2}\sum_i K_b (db_i)^2 + \frac{1}{2}\sum_j K_\theta (d\theta_j)^2 , \qquad (1)$$

where $db_i$ is the elongation of bond $i$ and $d\theta_j$ is the variance of the bond angle $j$. $K_b$ and $K_\theta$ are the corresponding force constants.

To elucidate the difference between the SSM and the beam model, we analyze the relation of the two models in armchair and zigzag graphene nanoribbons (see Fig. 1) under different loading conditions.

*2.1. The comparison between stick-spiral and beam models under the coupling force and moment*

Most researchers calculated the elastic properties of CNTs and graphene nanoribbons under different loading conditions with beam models using a constant BBS ($EI/b=K_\theta$ or $EI/b=0.5K_\theta$) (Li and Chou, 2003, 2004; Kasti, 2007; Li and Guo 2008); $E$, $I$ and $b$ are the Young's modulus, the moment of inertia and the initial bond length of the beam. Based on the energy equivalence between the SSM and the beam models, we find that the BBS in armchair and zigzag graphene nanoribbons under uniaxial tension and pure shear is $EI/b=0.5K_\theta$. For the finite width armchair graphene sheet under coupling loading force $F$ and moment $M$ ($\alpha=\beta$ and $b_1=b_2=a=b$ here), the BBS $EI/b$ should be employed, see Fig. 2. It should be noted that only the in-plane bending is considered in this paper.

For the SSM, the force and the moment equilibrium lead to

$$\begin{cases} F_1 = K_b db_1 \\ F_2 \cos\left(\dfrac{\alpha}{2}\right) = K_b db_2 \\ M = K_\theta \left(d\beta_1 + d\beta_2\right) \\ \dfrac{M}{2} + F_2 \sin\left(\dfrac{\alpha}{2}\right)b = K_\theta d\beta_1 + K_\theta \left(d\beta_1 - d\beta_2\right) \\ \dfrac{M}{2} - F_2 \sin\left(\dfrac{\alpha}{2}\right)b = K_\theta d\beta_2 - K_\theta \left(d\beta_1 - d\beta_2\right) \end{cases}, \qquad (2)$$

where $b_1=b_2=b$, $F_1=2F_2$, $d\beta_1$ and $d\beta_2$ are the angle increments.

The total energy of the stick $U_T$ can be written as

$$\begin{aligned}
U_T &= U_b + U_\theta \\
&= \frac{1}{2}K_b(db_1)^2 + 2\times\frac{1}{2}K_b(db_2)^2 + 2\times\left(\frac{1}{2}K_\theta(d\beta_1)^2 + \frac{1}{2}K_\theta(d\beta_2)^2 + \frac{1}{2}K_\theta(d\beta_1-d\beta_2)^2\right) \\
&= \frac{1}{2}K_b(db_1)^2 + K_b(db_2)^2 + 2K_\theta(d\beta_1)^2 + 2K_\theta(d\beta_2)^2 - 2K_\theta d\beta_1 d\beta_2
\end{aligned} \qquad (3)$$

The total energy of the beam model $U_{\text{Tbeam}}$ can be written as

$$U_{Tbeam} = U_{Fbeam} + U_{Mbeam}$$

$$= \int_0^b \frac{\left(F_1 \cos\left(\frac{\alpha}{2}\right)\right)^2}{2EA} dx_b + 2\int_0^b \frac{\left(F_2 \cos\left(\frac{\alpha}{2}\right)\right)^2}{2EA} dx_b + \int_0^b \frac{\left(\frac{M}{2} - F_2 \sin\left(\frac{\alpha}{2}\right) x_b\right)^2}{2EI} dx_b$$

$$+ \int_0^b \frac{\left(\frac{M}{2} + F_2 \sin\left(\frac{\alpha}{2}\right) x_b\right)^2}{2EI} dx_b + \int_0^b \frac{M^2}{2EI} dx_b$$

$$= \frac{\left(F_1 \cos\left(\frac{\alpha}{2}\right)\right)^2 b}{2EA} + \frac{\left(F_2 \cos\left(\frac{\alpha}{2}\right)\right)^2 b}{EA} + \frac{3M^2 b}{4EI} + \frac{\left(F_2 \sin\left(\frac{\alpha}{2}\right)\right)^2 b^3}{3EI} \quad , \quad (4)$$

where $x_b$ is the local coordinate systems along the beam, $A$ is the cross section area of the beam, and $U_{Fbeam}$ and $U_{Mbeam}$ are the strain energy from the force and moment, respectively.

Let $U_{Tbeam} = U_T$, then the BBS can be obtained from Eq. (2)-Eq. (4)

$$\frac{EI}{b} = K_\theta \frac{9 + 4\left(\frac{N}{M}\right)^2}{6 + 8\left(\frac{N}{M}\right)^2} \quad , \quad (5)$$

where $N = 1/2 F_1 \sin(\alpha/2) b$.

Similarly, we obtain the value of $EI/b$ when the beam model of the finite width zigzag nanoribbons in Fig. 1b is under the coupling force and moment (see Figs. 2 c and d):

$$\frac{EI}{b} = \frac{K_\theta}{2} \cdot \frac{\left(F \cos\frac{\alpha}{2} b\right)^2 + 3M^2 - 3MF \cos\frac{\alpha}{2} b}{\left(M - F \cos\left(\frac{\alpha}{2}\right) b\right)^2} = \frac{K_\theta}{2} \cdot \frac{3 + \left(\frac{N}{M}\right)^2 - 3\left(\frac{N}{M}\right)}{\left(1 - \frac{N}{M}\right)^2} \quad , \quad (6)$$

where $N = F \cos(\alpha/2) b$.

Comparing Eq. (5) with Eq. (6), the distributions of the BBS in the zigzag graphene nanoribbons are different to those in the armchair nanoribbons. Therefore, it is not suitable to use the same *EI/b* to calculate the corresponding mechanical properties under coupling loading force and moment in the finite width armchair and zigzag graphene nanoribbons.

The distribution of the BBS versus *N/M* in Eq. (5) is plotted in Fig. 3a. We find that the BBS strongly depends on the loading condition and is in the range of $0.5K_\theta \leq EI/b \leq 1.5K_\theta$ for the different loading conditions in the finite width armchair graphene nanoribbons. When *N/M* =0, Eq. (5) is degenerated into $EI/b=1.5K_\theta$ under pure moment *M* condition. When $N/M \to \infty$ (or $-\infty$), Eq. (5) is degenerated into $EI/b=0.5K_\theta$ under loading force *F* condition.

The distribution of the BBS versus *N/M* for the zigzag graphene nanoribbons is plotted in Fig. 3b. It also strongly depends on the loading condition. When *N/M* =0, $EI/b=1.5K_\theta$; when $N/M \to \infty$ (or $-\infty$), $EI/b=0.5K_\theta$; when $N/M \to 1$, $EI/b \to \infty$; when $N/M \to 3$, $EI/b \to 0.375K_\theta$ (the minimum).

In summary, the BBS of the SSM and the beam model differs and depends on the chirality and loading condition.

*2.2. The value of surface Young's modulus from stick-spiral and beam models*

In this section, we will compare the value of surface Young's modulus $Y_s$ ($E=Y_s/t=\sigma_s/(\varepsilon t)$) obtained from the SSM and the beam model; *E* and *t* denote the Young's modulus and thickness of the graphene sheet, and $\sigma_s$ is the surface stress which is equal to the stress multiplied by the thickness *t* of the graphene sheet (Chang and Gao, 2003). Moreover, we derive the expressions of $Y_s$ under uniaxial tension in armchair and zigzag nanoribbons.

For the zigzag graphene sheet in Fig. 1b under a uniform tensile stress $f$ along $x$ direction (Chang and Gao, 2003), we define the strain as

$$\varepsilon = \frac{d\left(2b\sin\frac{\alpha}{2}\right)}{2b\sin\frac{\alpha}{2}} = db\left(\frac{1}{b} + \frac{b\cos^2\frac{\alpha}{2}K_b}{6K_\theta \sin^2\frac{\alpha}{2}}\right), \tag{7}$$

The surface Young's modulus $Y_s$ can be derived by the SSM (Chang and Gao, 2003)

$$Y_s = \frac{F}{\left(b + \frac{b}{2}\right)\varepsilon} = \frac{6K_\theta K_b b \sin\frac{\alpha}{2}}{\frac{3}{2}b\left(6K_\theta \sin^2\frac{\alpha}{2} + K_b b^2 \cos^2\frac{\alpha}{2}\right)} = \frac{8\sqrt{3}K_\theta K_b}{18K_\theta + K_b b^2}, \tag{8}$$

where $F = 3/2 fb$ (Li and Guo, 2008).

For the beam model, the elastic strain energy of the structure should be equal to the external work.

$$U_{work} = \frac{1}{2}F\Delta L = \frac{F^2 L}{2E'A'} = \frac{\left(\frac{3}{2}fb\right)^2 L}{2E'A'}, \tag{9}$$

$$U_{Tbeam} = \frac{\left(3fb\cos\left(\frac{\alpha}{2}\right)\sin\left(\frac{\alpha}{2}\right)\right)^2 b}{EA} + \frac{9\left(f\cos^2\left(\frac{\alpha}{2}\right)\right)^2 b^5}{192EI}. \tag{10}$$

where $L = 2b\sin(\alpha/2)$, $E' = E = Y_s/t$ and $A' = 1.5b$ in one cell of graphene sheets (Chang and Gao, 2003; Li and Guo, 2008). Defining $EA/b = K_b$, and using Eq. (5), Eq. (6) and $U_{work} = U_{Tbeam}$, we obtain $Y_s$

$$Y_s = \frac{6K_\theta K_b b \sin\frac{\alpha}{2}}{\frac{3}{2}b\left(6K_\theta \sin^2\frac{\alpha}{2} + K_b b^2 \cos^2\frac{\alpha}{2}\right)} = \frac{8\sqrt{3}K_\theta K_b}{18K_\theta + K_b b^2}. \tag{11}$$

Note that Eq. (8) for the SSM and Eq. (11) of the beam model are identical.

We now compare the results of the two models with results from MD simulation. The value of $Y_s$ from the two models with different $EI/b$ is plotted in Fig. 4. When $K_b$=742 nN/nm, $K_\theta$=1.42 nN nm and $\alpha$=120° (Chang and Gao, 2003), the value of $Y_s$ is equal to 360 GPa nm in Eq. (8) and Eq. (11) which is in very good agreement with the MD result $Y_s$=350±20 GPa nm from Sanchez-Portal et al. (1999) and Van Lier et al. (2000). The results for $Y_s$ depending on the BBS are also plotted in Fig. 4. The values of $Y_s$ range from 321 GPa nm to 571 GPa nm under coupling force and moment. Those results are quite different to the MD result, when we use the BBS of the zigzag graphene sheet in Eq. (6). For example, the value of $Y_s$ (478 GPa nm) is about 1.33 times of the MD result using $EI/b$=1.5$K_\theta$ under pure bending condition ($N/M$→0). The value of $Y_s$ (360 GPa nm) is identical with MD result for $EI/b$=0.5$K_\theta$ under loading force ($N/M$→∞ or -∞) or uniaxial tension conditions. When $N/M$→1 in Eq. (6), $EI/b$→∞ leads to the maximum $Y_s$=571 GP nm which is about 1.59 times of the MD result. When $N/M$→3 in Eq. (6) (see Fig. 3b), $EI/b$=0.375$K_\theta$ leads to the minimum $Y_s$=321 GP nm which is about 0.89 time of the MD result. Therefore, it is crucial to give an exact force analysis in the structures so that the correct $EI/b$ can be obtained.

## 3. The validation using molecular dynamics simulation with harmonic potentials and finite element method

*3.1 Molecular dynamics simulation with harmonic potentials*

In this section, we present the results of FE and MD simulations with harmonic potentials. For the MD simulation, we keep the length $L$=14.7nm and the ratio $L/W$=1~60 in the armchair and $L/W$=1~52 in zigzag nanoribbons (see Fig. 5). Displacements are added at the left (green) and right (red) end layers. All MD simulations are performed using LAMMPS (Plimpton, 1995).

For uniaxial tension or pure shear, simulations are done at 0 $K$ and all atoms in the two end layers move 0.3 Å along the *x*- or *y*-direction at each time step, respectively, and every 0.5 bending degree at each time step for pure bending except for armchair $L/W$=60 (every 0.15 bending degree at each time in view of the large fluctuation). Afterwards, the structure is optimized for each displacement increment and the optimized structure is taken as the initial geometry for the next calculations. The energy minimization is performed using the conjugate-gradient method. A tolerance of relative energies between minimization iterations is set at 0.0 with a force tolerance of $10^{-8}$ to ensure a sufficiently minimized system. To model the bending deformation, rigid body translation is applied to the atoms in both end layers of the graphene sheets (see the green and red parts in Fig. 5), such that both end sections remain straight and are kept perpendicular to the deformed axis in each displacement increment (Iijima et al., 1996; Cao and Chen, 2006). The length of the middle line along the deformed axis in the graphene sheet remains unchanged and its curvature is uniform throughout the deformation.

First, we consider the armchair and zigzag graphene nanoribbons under uniaxial tension. The harmonic bond and angle potentials parameters $K_b$=742nN/nm and $K_\theta$=1.42 nN nm are adopted from Chang and Gao (Chang and Gao, 2003). The Lennard-Jones (LJ) pair potential $U_{LJ}$ between carbon and carbon is adopted as $U_{LJ} = 4\in\left[\left(\frac{\sigma}{r}\right)^{12} - \left(\frac{\sigma}{r}\right)^{6}\right]$ (Chang, 2007; Li and Guo, 2008), where *r* is the distance between the interacting atoms, $\in$ the depth of the potential, and $\sigma$ a parameter that is determined by the equilibrium distance. We use $\sigma$=3.407 Å and $\in =4.7483\times11.8^{-22}$ J (Kolmogorov et al., 2004; Vodenitcharova and Zhang, 2004; He et al., 2005; Chang, 2007).

In our MD simulations, the stress method and energy method are both used to calculate the Young's modulus and shear modulus. For the stress method, the stress on the surface of graphene sheet can be given by the component of the virial stress (Zhao et al., 2010)

$$\sigma_{ij} = -\frac{1}{V}\left(\sum_{i}^{N_{carbon}} m v_i v_i + \sum_{i=1, j<i}^{N_{carbon}} r_{ij}\frac{\partial U_{ij}}{\partial r_{ij}}\right), \qquad (12)$$

where $V$ is the current volume of the graphene sheet, $m_i$ is the mass of atom $i$, $v_i$ is the velocity, $r_{ij}$ is the displacement vector between the atoms $i$ and $j$, and $U_{ij}$ is the potential energy between atoms $i$ and $j$.

The idea for the energy method is that the increment of the total energy should be equal to the external work (Zhao et al., 2009b). The equation can be written as

$$\begin{cases}\sigma_c = \dfrac{1}{V_0}\dfrac{\partial U}{\partial \varepsilon} \\ M = \dfrac{1}{V_0}\dfrac{\partial U}{\partial \phi}\end{cases}, \qquad (13)$$

where $U$ and $\varepsilon$ are the total energy increment and tensile strain, $\sigma_c$ and $M$ are the tensile stress and bending moment on the left or right regions in Fig. 5, and $V_0$ and $\phi$ are the initial volume and bending angle.

The total energy for different tensile and shear strains are plotted in Fig. 7a. The surface tensile or shear stresses obtained from Eq. (13) are plotted in Fig.7b. Note that the surface stress is the stress multiplied by the thickness $t$ of the graphene sheet. Defining the surface tensile stress and the surface shear stress as $\sigma_s$ and $\tau_s$, the surface Young's modulus $Y_s$ and shear modulus $G$ is expressed as $Y_s=\sigma_s/\varepsilon$ and $G=\tau_s/\gamma$, where $\varepsilon$ and $\gamma$ are the tensile strain and shear strain. Fig. 7b shows that the difference of the surface stress-strain curves between the armchair and zigzag nanoribbons is very small. Those observations agree well with the results in the literature (Chang and Gao, 2003; Kasti, 2007; Li and Guo, 2008). Fig. 7b also shows that the surface stress-strain curves of the stress method are in very good agreement with those of the energy method. Our MD results agree with those of the available analytical models (see Fig. 7b) (Chang and Gao, 2003; Li and Guo, 2008). The energy method is adopted to obtain all MD results in the following text.

*3.2 Finite element method based on the beam model*

The FE beam structures of graphene sheets can be easily built from the coordinates of the graphene MD models (Fig. 5).

We adopt the stiffness $EA/b=K_b$ and $EI/b=0.5K_\theta$ with Young's modulus $E$=9.18~14.77TPa and Poisson's ratio $v$=0~0.4 from Li and Guo (2008). All the present FE calculations are performed using the commercial ANSYS 12.0 package with 2-node BEAM188 element.

The surface stress-strain curves along different directions for $E$=14.77TPa, $v$=0.1 and $EI/b=0.5K_\theta$ are plotted in Fig. 8. The difference of the stress-strain curves between the armchair and zigzag graphene nanoribbons are very small, which agrees well with the observations from Li and Guo (2008) and Sakhaee-Pour (2009).

In view of so small difference, we only study the effect of the Poisson's ratio $v$ on the stress-strain curves for the armchair sheet in Fig. 9. The surface stress-strain curves don't change with $v$ at all in Fig. 9, which means that the Young's modulus and shear modulus of graphene sheet are both independent of Poisson's ratio $v$ of the beam. Therefore, there is no limitation to use

Poisson's ratio $v$ (as $v=0\sim0.4$) of the beam element so that we can obtain the same Young's modulus and shear modulus. The Poisson's ratio $v=0.1$ are adopted in the following FE results.

As shown in Fig. 10, the effect of the beam Young's modulus on the surface stress-strain curves is also very small. Li and Guo's results are between the present two curves although $K_b$=723nN/nm and $K_\theta$=1.36 nN nm is chosen in their literature (Li and Guo, 2008).

Since the Poisson's ratio $v$ and the Young's modulus $E$ of the beam model have almost no effect on the elastic properties of the graphene nanoribbons, we choose $E$=14.77TPa and $v$=0.1 in all the following FE calculations.

*3.3 Results and discussion*

Fig. 11a plots the surface tensile stress ratios between the MD simulations with harmonic potentials and FE results based on the beam model under uniaxial tension. All the ratios are close to 1 for different $L/W$ in the armchair and the zigzag graphene nanoribbons. It means that the BBS, $EI/b=0.5K_\theta$ is correct to describe the elastic properties of graphene nanoribbons under tension and shear, which validates our analytical results in Section 2.1. Figs. 11b and c show that the bending moment ratios between the MD and FE results for the BBS of $EI/b=0.5K_\theta$. The ratios $M_{MD}/M_{FE}$ increase with decreasing width $W$ in both armchair and zigzag nanoribbons. The maximum ratios reach values up to 2.5 in the armchair nanoribbons and 1.25 in the zigzag nanoribbons. It indicates that the loading-mode dependent BBS in the armchair nanoribbons is more pronounced than the BBS in the zigzag nanoribbons.

We futher study the change of the corresponding bonds and angles in the armchair and zigzag graphene nanoribbons with different bending angles. The distributions of the bond length and the bond angles in the upper and lower regions are symmetric with regard to the middle line along the deformed axis. For the narrow sheets in Figs. 12a, d and Figs. 13a, d, the bond length and the angles change weakly with increasing bending angle. With increasing width, the bond length and the bond angles from the middle line to the free surface along the undeformed axis increase sharply with increasing bending angle, as shown in Figs. 12c, f and Figs. 13c, f.

For all sheets, the bond length and the bond angles in the middle regions change weakly, while the bond length in the upper and the lower regions is sharply elongated and shortened, respectively. Fig. 13d indicates that all the bond length change considerably in the narrow zigzag sheet, while the bond length does not change in the narrow armchair sheet. From the armchair sheets in Fig. 5, we find that the bonds $b_k$ ($k$=1, 2, 3, 01, 02, ⋯,08) are parellel to the deformed axial. When the armchair sheets are under pure bending, the bonds $b_k$ in the middle parts of the sheets are always under pure bending. Therefore, the ratio $m=N_b/N_{tc}$ increases with decreasing width, where $N_b$ is the number of the bonds subjected to bending and $N_{tc}$ is the number of the bonds subjected to tension/compression in the sheets. However, all of the bonds $c_l$ ($l$=0, 1, 2, 3, 01, 02, ⋯, 09, 010) in the zigzag sheets are not parallel to the deformed aixal. The ratio in the zigzag sheets is always less than that in the armchair sheets for the same value of *L/W*. It is the main reason that the bending moment ratios in the armchair sheets (see Figs. 11b and c) are larger than those in the zigzag sheets. The loading-mode dependent BBS of the armchair nanoribbons is more pronounced than that of the zigzag nanoribbons.

Fig. 14 and Fig. 15 show the spatial distributions of the bond length and the average angle variation in different armchair and zigzag graphene nanoribbons for a bending angle of 15 degree. The average angle variation of atom *i* is calculated by

$$\Delta\theta_i = \frac{1}{3}\sum_{i=1}^{3}|\theta_i - \theta_0|, \tag{14}$$

where $\theta_i$ ($i$=1,2,3) are the three angles around the atom $i$ at a given bending angle, and $\theta_0$ is the initial angle of 120 degrees.

The bond length and the average angle variation change weakly in the middle regions and sharply in the upper and lower regions. It indicates that the middle regions of all graphene nanoribbons are always subjected to bending, while the upper and lower regions are mainly under tension or compression, respectively. With increasing width, tension and compression dominate the bending properties of the graphene nanoribbons (see Figs. 14c and f or Figs. 15c and f). Conversely, with decreasing width, bending or coupling tension/compression-bending dominate the bending properties of the nanoribbons (see Figs. 14a and d or Figs. 15a and d). From our analysis in Section 2.1, the BBS $EI/b$=0.5$K_\theta$ should be used in uniaxial tension/compression/shear, while BBS $EI/b$=1.5$K_\theta$ in pure bending should be used for considerably narrow graphene nanoribbons. Therefore, it is reasonable to adopt the BBS 0.5$K_\theta$≤$EI/b$≤1.5$K_\theta$ and 0.375$K_\theta$≤$EI/b$ in the finite width armchair sheets and zigzag sheets under pure bending in Figs. 3a and b, respectively.

Fig. 16 illustrates the bending moment ratios for graphene nanoribbons for different BBS $EI/b$. The ratio is close to 1 when $EI/b$=1.5$K_\theta$ is used in our FE calculation with $L/W$=60 in Fig. 5a, which perfectly validates our analytical result in Section 2.1. Furthermore, $EI/b$=$K_\theta$ can be used to describe the elastic properties in $L/W$=30 (armchair) and $L/W$=52 (zigzag) graphene nanoribbons considering the domination of the coupling tensile/compressive-bending mode.

The BBS $EI/b$=0.5$K_\theta$ of the beam models describe the elastic properties well under uniaxial tension or pure shear. However, the BBS strongly depends on the width and the chirality of the graphene nanoribbons under pure bending or tensile-bending modes. When the width of the armchair graphene sheets becomes small enough ($L/W$=60), $EI/b$=1.5$K_\theta$ describes the bending behavior excellently under pure bending. With increasing width, $0.5K_\theta \leq EI/b \leq 1.5K_\theta$ and $0.375K_\theta \leq EI/b$ should be used to effectively describe the mechanical behavior in armchair and zigzag sheets, respectively.

In view of the extremely narrow structure of a single polyethylene PE chain, we further analyzed the elastic properties of the PE chain under different loading conditions too.

*3.4 The two models in a single polyethylene chain*

In this section, we study one PE chain under coupling loading force *f* and moment *m,* see Fig. 17a. In analogy to our analysis in Section 2.1, the value of the BBS *EI/b* of the PE can be written as

$$\frac{EI}{b} = \frac{K_{p\theta}}{3} \frac{3m^2 + \left(f\cos\frac{\theta}{2}b\right)^2 - 3mf\cos\frac{\theta}{2}b}{\left(m - f\cos\frac{\theta}{2}b\right)^2} = \frac{K_{p\theta}}{3} \frac{3 + \left(\frac{n}{m}\right)^2 - 3\left(\frac{n}{m}\right)}{\left(1 - \frac{n}{m}\right)^2}, \quad (15)$$

where $n=f\cos(\alpha/2)b$, *b* and *θ* are the initial bond length and angle of the PE chain, respectively, and $K_{p\theta}$ is the bond bending stiffness of PE (Zhao et al., 2011).

Eq. (15) and Eq. (6) differ only in the coefficients. The distribution of the bending stiffness in Eq. (15) over *n/m* is shown in Fig. 17b. The bending stiffness *EI/b* is larger than $0.25K_\theta$ for the different loading conditions. When *n/m*=0, Eq. (15) is degenerated into $EI/b=K_\theta$ under pure moment *m* condition. When *n/m*→∞ (or -∞), Eq. (15) is degenerated into $EI/b=K_\theta/3$ under loading force *F* condition. As *n/m*→1, Eq. (15) leads to *EI/b*→∞. The minimum $EI/b=0.25K_\theta$ is obtained for *n/m*→3.

To further validate the analytical results, we carried out the united-atom MD simulation and FE simulation in Fig. 18. The PE chain consists of 19 united-atom beads with a length $L$=2.28 nm. In the united atom approximation, the methyl groups (CH$_2$) are represented by a single "atom" and the effect of the hydrogen atoms on the polymer's configuration is accounted for in the potentials (Waheed, 2005; Shepherd, 2006; Zhao et al., 2011). The parameters of the harmonic potentials are $K_b$=700 Kcal/mol Å$^2$, $K_\theta$=120 Kcal/mol, $b$=1.53 Å, $\theta$=109.5°. The LJ pair potential (see Section 3.1) with $\in$=0.112 Kcal/mol and $\sigma$=4.01 Å is adopted (Waheed, 2005; Shepherd, 2006; Zhao et al., 2011).

Fig. 18a compares the tensile stress-strain curves of the united-atom model with the FE model. The Young's moduli $Y_{UA}$ for both models are in excellent agreement. A cross sectional area of 17.3 Å$^2$ is adopted in the FE model which is equal to the average area of one PE chain in crystalline PE (Henry and Chen, 2008; Zhao et al., 2011; Jiang et al., 2012). The Young's moduli $Y_{UA}$=190.4 GPa and $Y_{FE}$=192.6 GPa are obtained by fitting the data in the range of 10% tensile stress-strain curves in Fig. 18a. Those results are in good agreement with the analytical value 195.1 GPa of crystalline PE in our previous work (Zhao et al., 2011). The distribution of the bending moment ratios between the united-atom and FE models for $EI/b=1/3K_\theta$ and $EI/b=K_\theta$ versus bending angles are plotted in Fig. 18b. The bending moment ratios between those models at $EI/b=1/3K_\theta$ are always higher than 2.28, while the ratios are close to 1 when $EI/b=K_\theta$. The result effectively validates Eq. (15) under tension and pure bending conditions.

The above analysis shows that the difference between the stick-spiral and the beam models is independent of the materials because all the parameters of the beam model are obtained from the harmonic potential.

Moreover, one has to be taken when the beam model is employed for the crystalline (or amorphous) polymers or other biopolymers (Zhao et al., 2010; Zhao et al., 2011; Zhao et al., 2012; Zheng and Sept, 2008), as their structures are composed of many single molecular chains and there are only weak van der Waals and coulomb interactions (Zhao et al., 2011) between two chains. It is possible to observe more pronounced difference between the MD and FE results in large deformation under uniaxial tension and pure bending (see Fig. 18b) if one uses a same constant $EI/b=1/3K_\theta$.

Despite of the difference between the SSM and the beam model, it is not clear yet which model is better suitable to predict the elastic properties of carbon nanotubes and graphene sheets.

Therefore, we carried out additional MD simulation with the AIREBO potential (Plimpton, 1995), which is commonly used to obtain the mechanical properties of graphene nanoribbons (Zhao et al., 2009a).

## 4. The comparison of the two models with molecular dynamics simulation of AIREBO potential

We adopt the setup from Section 3.1 but use the AIREBO potential in this section (Zhao et al., 2009a). The total energy increments with the harmonic potentials and the AIREBO potential under uniaxial tension and pure bending are plotted in Fig. 19 and Fig. 20, respectively. Higher values are obtained for the harmonic potentials.

Fig. 21 shows the elastic properties of the different models, $C_A$, $C_H$ and $C_{FE}$ are the stretching stiffness (the bending stiffness of a total nanoribbon) of the AIREBO, the harmonic and the FE results, respectively, and $D_A$, $D_H$ and $D_{FE}$ are the bending stiffness (in-plane bending stiffness of each nanoribbon) of the AIREBO, the harmonic and the FE results, respectively. For all the FE results, we used $EI/b=0.5K_\theta$. Modeling each nanoribbon as a beam under small deformation condition, the stretching stiffness $C$ and bending stiffness $D$ per unit volume from Eq. (13) can be written as

$$C = Y_g = 2U_{tension}/(V_0\varepsilon^2), \text{ under uniaxial tension,} \tag{16}$$

$$D = 2U_{bending}/(V_0\theta^2), \text{ under pure bending,} \tag{17}$$

where $U_{tension}$ and $U_{bending}$ are the total tension energy increment and the bending energy increment, $V_0$ is the initial volume, $Y_g$ is the Young's modulus, $\varepsilon$ is the tensile strain and $\theta$ is the bending angle of each graphene nanoribbon.

From Eq. (16) and Eq. (17), the stiffnesses $C$ and $D$ for different $L/W$ can be obtained by fitting the data in Fig. 19 and Fig. 20, in which the data in the range of 0~6% tensile strain and 0~10 degrees of bending angle are used in the fitting procedure. Fig. 21 shows that the values of $C_H/C_A$ are about 1.26~1.3 and 0.99~1.1 in different width zigzag and armchair graphene nanoribbons. The values of $C_{FE}/C_A$ (about 1.29~1.35 and 0.84~1.14 in zigzag and armchair nanoribbons) are similar to those of $C_H/C_A$.

Under pure bending condition, $D_H/D_A$ (from 1.18~1.24) and $D_{FE}/D_A$ (from 1.16~1.22) are almost identical in the finite width zigzag nanoribbons except for $L/W$=52 ($D_H/D_A$=1.36, $D_{FE}/D_A$=1.08). All values of $D_H/D_A$ and $D_{FE}/D_A$ are very close to the values of $C_H/C_A$ and $C_{FE}/C_A$ besides the value of $D_{FE}/D_A$ at $L/W$=52 which is a little lower. In other words, the BBS in the zigzag nanoribbons is insensitive to different loading modes except for the ultra-narrow nanoribbon with $L/W$=52. A similar phenomenon can be observed from Fig. 11.

For the armchair nanoribbons, the results of $D_H/D_A$ (1.11~1.85) and $D_{FE}/D_A$ (0.7~1.25) are much higher and lower than those of $C_H/C_A$ (0.99~1.1) and $C_{FE}/C_A$ (0.84~1.14) with increasing $L/W$, respectively. Hence, the SSM overestimates the values, while the beam model underestimates the values. Therefore, we suggest to choose the average value between the SSM and beam models in the narrow graphene nanoribbons under pure bending. Above analysis indicates that the loading-mode dependent BBS in the armchair nanoribbons is more pronounced than that in the zigzag nanoribbons.

## 5. Concluding remarks

We extensive studied the difference between the stick-spiral and beam models in the finite width armchair and zigzag graphene nanoribbons and the single PE chain. Based on the total energy equilibrium in the two models, the closed-form expressions of the BBS are derived under uniform tension, pure shear, pure bending, loading force, coupling force and bending conditions.

By comparisons of the two models, we found that the BBS of the beam model strongly depends on the loading modes in narrow graphene nanoribbons. Based on the MD simulations with harmonic potentials and FE results, the BBS $EI/b=0.5K_\theta$ of the beam model can be used to describe the elastic properties well under uniaxial tension or pure shear. Under pure bending or coupling tensile-bending modes, the BBS depends on the width and chirality of the graphene nanoribbons. When the width of the armchair graphene sheets becomes small enough, $EI/b=1.5K_\theta$ can be used to describe the bending behavior effectively under pure bending. With increasing width, $0.5K_\theta \leq EI/b \leq 1.5K_\theta$ and $0.375K_\theta \leq EI/b$ should be used to model the mechanical behavior in the armchair and the zigzag sheets, respectively. For a single PE chain, similar phenomena can be found, in which $1/3K_\theta \leq EI/b \leq K_\theta$ under different loading conditions.

We also found that the difference of the stick-spiral and the beam models exists and they are independent of the materials because all parameters of the beam model are obtained from the harmonic potentials. For the narrow graphene nanoribbons or a single PE chain, the maximum difference can exceed 300% in different loading modes, while the difference is completely concealed in higher width nanoribbons.

Therefore, the beam model should be used carefully to model crystalline polymers and biomaterials in view of van der Waals and coulomb interactions between any two chains. It is possible to obtain more pronounced difference between the MD results with harmonic potentials and FE results in large deformation under uniaxial tension and pure bending if one uses the same constant $EI/b=1/3K_\theta$ in a single PE chain or $EI/b=0.5K_\theta$ in narrow armchair graphene nanoribbons, respectively.

When the results of the MD models with harmonic potentials and the FE calculation based on the beam model are compared with those of the MD results with the AIREBO potential, the SSM overestimates and the beam model underestimates the values of the armchair nanoribbons under pure bending condition, respectively.

**Acknowledgements: inancial support for** the work from the Germany Science Foundation (DFG) project. The authors would like to thank Prof. T.C. Chang and Dr. M. Silani for many useful discussions.

**References:**

Badia, S., Bochev, P., Lehoucq, R., Parks, M. L., Fish, J., Nuggehally, M., Gunzburger M., 2007. A force-based blending model for atomistic-to-continuum coupling. Int. J. Multiscale Comput. Eng. 5, 387–406.

Bian, J.J, Wang, G.F., 2012. A multi-scale approach of amorphous polymer from coarse grain to finite element. Comput. Mater. Sci. 57, 8-13.

Capaldi, F.M., Boyce, M.C., Rutledge, G.C., 2004. Molecular response of a glassy polymer to active deformation. Polymer 45, 1391.

Chang, T., Gao, H., 2003. Size-dependent elastic properties of a single-walled carbon nanotube via a molecular mechanics model. J. Mech. Phys. Solids 51, 1059-1074.

Chang, T., Geng, J., Guo, X., 2005. Chirality- and size-dependent elastic properties of single-walled carbon nanotubes. Appl. Phys. Lett. 87, 251929.

Chang, T., 2007. Torsional behavior of chiral single-walled carbon nanotubes is loading direction dependent. Appl. Phys. Lett. 90, 201910.

Chang, T., 2010. A molecular based anisotropic shell model for single-walled carbon nanotubes. J. Mech. Phys. Solids 58, 1422-1433.

Chopra, N.G., Luyken, R.J., Cherrey, K., Crespi, V.H., Cohen, M.L., Louie, S.G., Zettl, A., 1995. Boron nitride nanotubes. Science 269, 966-967.

Cao, G., Chen, X., 2006. Bulking of single-walled carbon nanotubes upon bending: Molecular dynamics simulations and finite element method. Phys. Rev. B 73, 155435.

Di Matteo, A., Müller-Plathe, F., Milano, G., 2007. From Mesoscale back to Atomistic Models: A Fast Reverse-Mapping Procedure for Vinyl Polymer Chains. J. Phys. Chem. B 111, 2765-2773.

Hageman, J.C.L., Meier, R.J., Heinemann, M., Groot, R.A., 1997. Young's modulus of crystalline polyethylene from ab initio molecular dynamics. Macromolecules 30, 5953-5957.

He, X.Q., Kitipornchai, S., Liew, K.M., 2005. Buckling analysis of multi-walled carbon nanotubes: a continuum model accounting for van der Waals interaction. J. Mech. Phys. Solids 53, 303-326.

He, X.Q., Kitipornchai, S., Wang, C.M., Liew, K.M., 2005. Modeling of van der Waals force for infinitesimal deformation of multi-walled carbon nanotubes treated as cylindrical shells. Int. J. Solids Struct. 42, 6032~6047.

Hernandez, E., Goze, C., Bernier, P., Rubio, A., 1998. Elastic properties of C and $B_xC_yN_z$ composite nanotubes. Phys. Rev. Lett. 80, 4502-4505.

Henry, A., Chen, G., 2008. High thermal conductivity of single polyethylene chains using molecular dynamics simulations. Phys. Rev. Lett. 101, 235502.

Iijima, S., Brabec, C., Maiti, A., Bernholc, J., 1996. Structural flexibility of carbon nanotubes. J. Chem. Phys. 104, 2089-2092.

Jiang, J., Wang, J.S., Li B., 2009. Young's modulus of graphene: a molecular dynamics study. Phys. Rev. B 80, 113405.

Jiang, J., Zhao, J., Kun, Z., Rabczuk, T., 2012. Superior thermal conductivity and extremely high mechanical strength in polyethylene chains from ab initio calculation. J. Appl. Phys., 111, 124304.

Jiang, L., Guo, W., 2011. A molecular mechanics study on size-dependent elastic properties of single-walled boron nitride nanotubes. J. Mech. Phys. Solids 59, 1204-1213.

Kasti, N., 2007. Zigzag carbon nanotubes-Molecular/structural mechanics and finite element method. Int. J. Solids Struct. 44, 6914-6929.

Kolmogorov, A.N., Crespi, V.H., Schleier-Smith, M.H., Ellenbogen, J.C., 2004. Nanotube-Substrate Interactions: Distinguishing Carbon Nanotubes by the Helical Angle. Phys. Rev. Lett. 92, 085503.

Li, C.Y., Chou, T.W., 2003. A structural mechanics approach for the analysis of carbon nanotubes. Int. J. Solids Struct. 40, 2487-2499.

Li, C.Y., Chou, T.W., 2004. Elastic properties of single-walled carbon nanotubes in transverse directions. Phys. Rev. B 69, 073401.

Li, H., Guo, W., 2006. Finite element model with equivalent beam elements of single-walled carbon nanotubes. Acta Mech. Sin. 38, 488-495.

Li, H., Guo, W., 2008. Transversely isotropic elastic properties of single-walled carbon nanotubes by a rectangular beam model for the C-C bonds. , J. Appl. Phys. 103, 103501.

Peng, J., Wu, J., Hwang, K.C., Song, J., Huang, Y., 2008. Can a single-wall carbon nanotube be modeled as a thin shell? J. Mech. Phys. Solids 56, 2213-2224.

Plimpton, S., 1995. Fast parallel algorithms for short-range molecular dynamics. J. Comput. Phys. 117, 1-19.

Ru, C.Q., 2000. Effect of van der Waals forces on axial buckling of a double-walled carbon nanotube. J. Appl. Phys. 87, 7227-7231.

Ru, C.Q., 2001. Axially compressed buckling of a doublewalled carbon nanotube embedded in an elastic medium. J. Mech. Phys. Solids 49, 1265-1279.

Sakhaee-Pour, A., 2009. Elastic properties of single-layered graphene sheet. Solid State Commu. 14, 91-95.

Sanchez-Portal, D., Artacho, E., Soler, J.M., Rubio, A., Ordejon, P., 1999. Ab-initio structural, elastic, and vibrational properties of carbon nanotubes. Phys. Rev. B 59, 12678.

Shepherd, J.E., 2006. Multiscale modeling of the deformation of semi-crystalline polymers. Ph. D thesis. Georgia Institute of Technology, USA.

To, C., 2006. Bending and shear moduli of single-walled carbon nanotubes. Finite Elem. Anal. Des. 42, 404-413.

Tserpes, K.I., Papanikos, P., 2005. Finite element modeling of single-walled carbon nanotubes. Composites B 36, 468-477.

Vaccarini, L., Goze, C., Henrard, L., Hernandez, E., Bernier, P., Rubio, A, 2000. Mechanical and electronic properties of carbon and boron-nitride nanotubes. Carbon 38, 1681-1690.

Vodenitcharova, L., Zhang L.C, 2004. Mechanism of bending with kinking of a single-walled carbon nanotube. Phys. Rev. B 69, 115410.

Waheed, N., 2005. Molecular simulation of crystal growth in alkane and polyethylene melts. Ph. D thesis. Cornell University, USA.

Wang, C.Y., Zhang, Y.Y., Wang, C.M., Tan, V.B.C., 2007. Buckling of carbon nanotubes: Aliterature survey. J. Nanosci. Nanotech. 7, 4221-4247.

Wu, J., Hwang, K.C., Huang, Y., 2008. An atomistic-based finite-deformation shell theory for single-wall carbon nanotubes. J. Mech. Phys. Solids 56, 279-292.

Xia, J.R., Gama, B.A., Gillespie Jr., J.W., 2005. An analytical molecular structural mechanics model for the mechanical properties of carbon nanotubes. Int. J. Solids Struct. 42, 3075-3092.

Yakobson, B.I., Brabec, C.J., Bernholc, J., 1996. Nanomechanics of carbon tubes: instability beyond linear response. Phys. Rev. Lett. 76, 2511-2514.

Zhang , Y.Y., Wang, C.M, Duan, W.H., Xiang, Y., Zong, Z., 2009. Assessment of continuum mechanics models in predicting buckling strains of single-walled carbon nanotubes. Nanotechnology 20, 395707.

Zhang, Z., Guo, W., Tai, G., 2007. Coaxial nanocable: Carbon nanotube core sheathed with boron nitride nanotube. Appl. Phys. Lett. 90, 133103.

Zhang, Z., Guo, W., 2008. Energy-gap modulation of BN ribbons by transverse electric fields: First-priciples calculations. Phys. Rev. B 77, 075403.

Zhao, H., Min, K., Aluru, R., 2009a. Size and chirality dependent elastic properties of graphene nanoribbons under uniaxial tension. Nano Lett., 9, 3012-3015.

Zhao, J., Zhou, B., Liu, B.G., Guo, W., 2009b. Elasticity of single-crystal calcite by first-principles calculations. J. Comput. Theor. Nanosci. 6, 1181-1188.

Zhao, J., Nagao, S., Zhang, Z.L., 2010. Thermo-mechanicyal properties dependence on chain length in bulk polyethylene: Coarse-grained molecular dynamics simulations, J. Mater. Res., 25, 537-544.

Zhao, J., Guo, W., Zhang, Z.L., Rabczuk, T., 2011. Size-dependent elastic properties of crystalline polymers via a molecular mechanics model. Appl. Phys. Lett. 99, 241902.

Zhao, J., Guo, W., Rabczuk, T., 2012. An analytical molecular mechanics model for the elastic properties of crystalline polyethylene. J. Appl. Phys. doi: 10.1063/1.4745035.

Zheng, X, Sept, D., 2008. Molecular modeling of the cytoskeleton. Methods Cell Biol. 84, 893-910.

## The Caption of the Figures

Fig. 1 The beam structures of the armchair and the zigzag graphene nanoribbons in the FE method based on the beam elements ($L/W$=1, $L$=14.7 nm).

Fig. 2. (a) One cell of a finite width armchair graphene sheet under coupling loading force $F$ and moment $M$, (b) angle increment of (a) for the stick-spiral model, (c) one cell of a finite width zigzag graphene sheet under coupling loading force $F$ and moment $M$, (d) angle increment of (c) for the stick-spiral model.

Fig. 3. The distribution of beam bending stiffness with $N/M$ under coupling loading force $F$ and moment $M$ in the finite width graphene nanoribbons, (a) armchair, (b) zigzag.

Fig. 4. The value of $Y_s$ from two models and different beam bending stiffness $EI/b$ in the finite width graphene nanoribbons.

Fig. 5 Finite width armchair and zigzag graphite nanoribbons under pure bending at bending angle=15 degree, (a) armchair $L/W$=60, (b) armchair $L/W$=20, (c) armchair $L/W$=7.5, (d) zigzag $L/W$=52, (e) zigzag $L/W$=20.8, (f) zigzag $L/W$=7.4.

Fig. 6 The zoomed-in view of the graphene nanoribbons in Fig. 5, (a) a zoomed-in view of Fig. 5a, (b) a zoomed-in view of Fig. 5b, (c) a zoomed-in view of Fig. 5c, (d) a zoomed-in view of Fig. 5d, (e) a zoomed-in view of Fig. 5e, (f) a zoomed-in view of Fig. 5f.

Fig. 7 (a) The total energy-strain and (b) the surface stress-strain curves of the armchair and the zigzag graphene sheet under uniaxial tension and pure shear in Fig. 1a and b.

Fig. 8 The surface tensile and shear stress-strain curves of FE method in Fig. 1.

Fig. 9 The surface stress-strain curves of the FE method in Fig. 1a.

Fig. 10 The surface tensile and shear stress-strain curves of the FE method with different beam Young's modulus in Fig. 10b along *x*-direction tension and *xy*-direction shear.

Fig. 11 The surface tensile stress ratios and bending moment ratios between MD and FE results in finite width armchair and zigzag graphene nanoribbons, (a) the surface tensile ratios in the armchair and the zigzag nanoribbons, (b) bending moment ratios in the armchair nanoribbons, (c) bending moment ratios in the zigzag nanoribbons.

Fig. 12 Bond length distributions of the armchair and zigzag graphene nanoribbons with different bending angles in Fig. 5, (a) armchair $L/W$=60, (b) armchair $L/W$=20, (c) armchair $L/W$=7.5, (d) zigzag $L/W$=52, (e) zigzag $L/W$=20.8, (f) zigzag $L/W$=7.4.

Fig. 13 Angle distributions of armchair and zigzag graphene nanoribbons with different bending angles in Fig. 5, (a) armchair $L/W$=60, (b) armchair $L/W$=20, (c) armchair $L/W$=7.5, (d) zigzag $L/W$=52, (e) zigzag $L/W$=20.8, (f) zigzag $L/W$=7.4.

Fig. 14 The spatial distributions of the bond length in armchair and zigzag graphene nanoribbons at the bending angle 15 degree, (a) armchair $L/W$=60, (b) armchair $L/W$=20, (c) armchair $L/W$=7.5, (d) zigzag $L/W$=52, (e) zigzag $L/W$=20.8, (f) zigzag $L/W$=7.4.

Fig. 15 The spatial distributions of the average angle increment in armchair and zigzag graphene nanoribbons at the bending angle 15 degree, (a) armchair $L/W$=60, (b) armchair $L/W$=20, (c) armchair $L/W$=7.5, (d) zigzag $L/W$=52, (e) zigzag $L/W$=20.8, (f) zigzag $L/W$=7.4.

Fig. 16 Bending moment ratios between MD and FE results for graphene nanoribbons with different $EI/b$.

Fig. 17. (a) One cell of a crystalline polyethylene chain under coupling loading force *f* and moment *m*, (b) the distribution of beam bending stiffness with $n/m$.

Fig. 18 (a) Stress-strain curves under tension and (b) bending moment ratios between united-atom MD and FE results in a single PE chain.

Fig. 19 Total energy increment with present harmonic potentials and AIREBO potential in armchair and zigzag graphene nanoribbons under tension.

Fig. 20 Total energy increment with present harmonic potentials and AIREBO potential in armchair and zigzag graphene nanoribbons under pure bending.

Fig. 21 Distribution of the two models to AIREBO ratios with $L/W$.

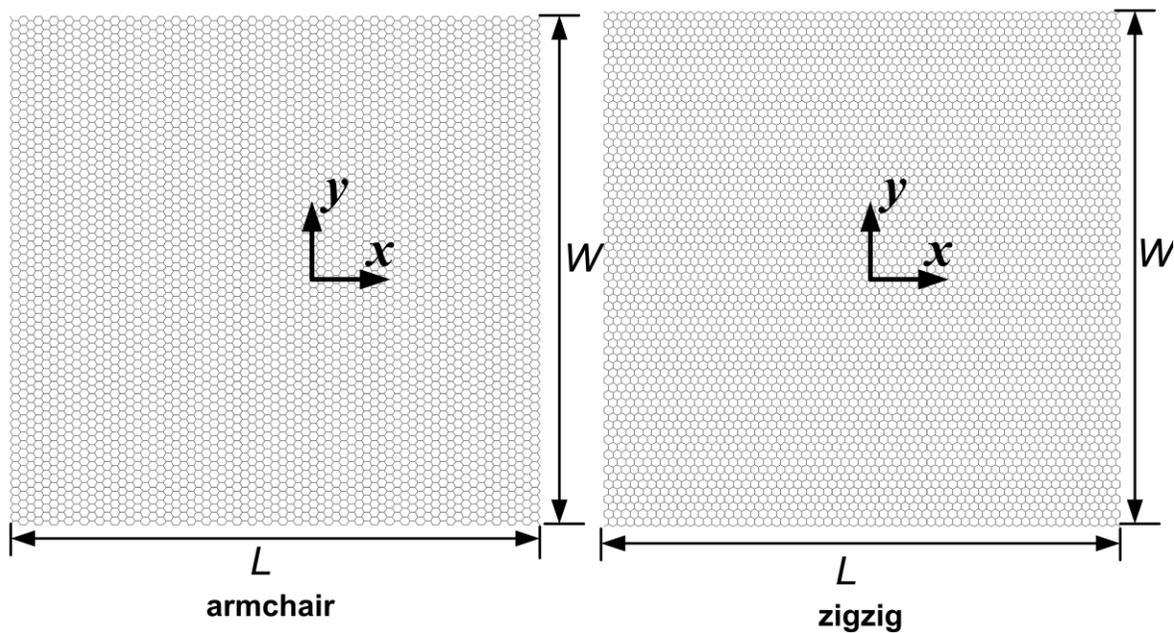

Fig. 1

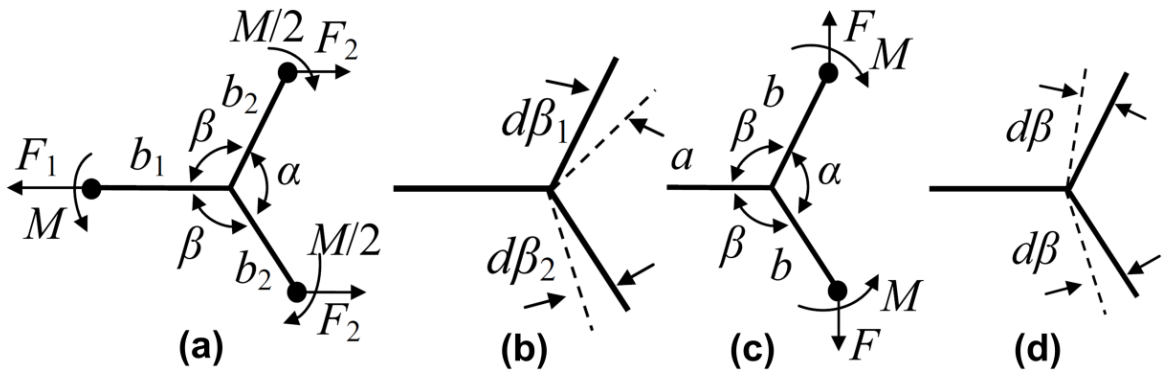

Fig. 2

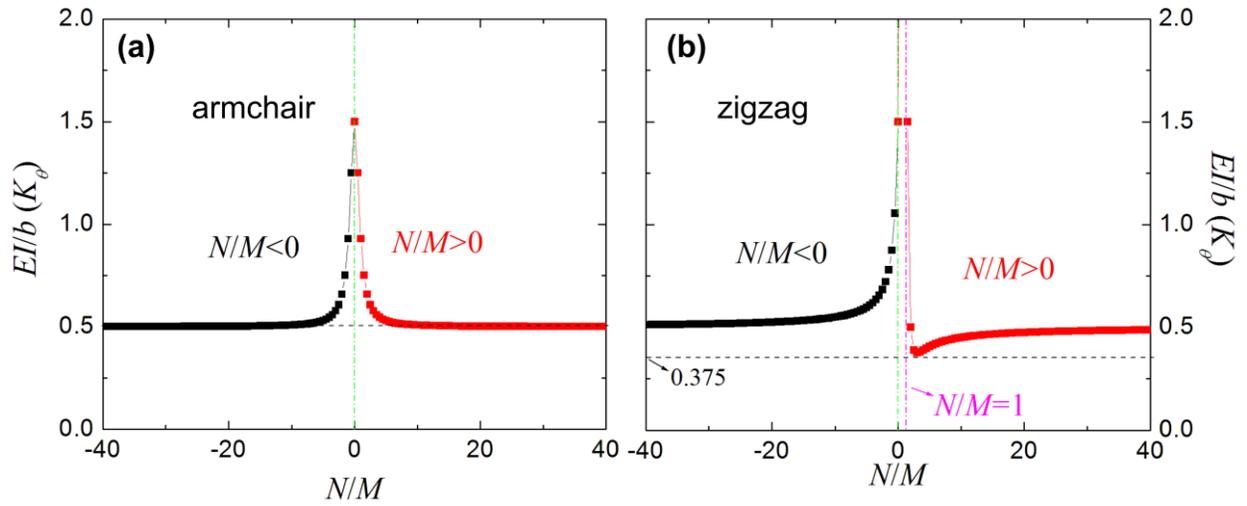

Fig. 3

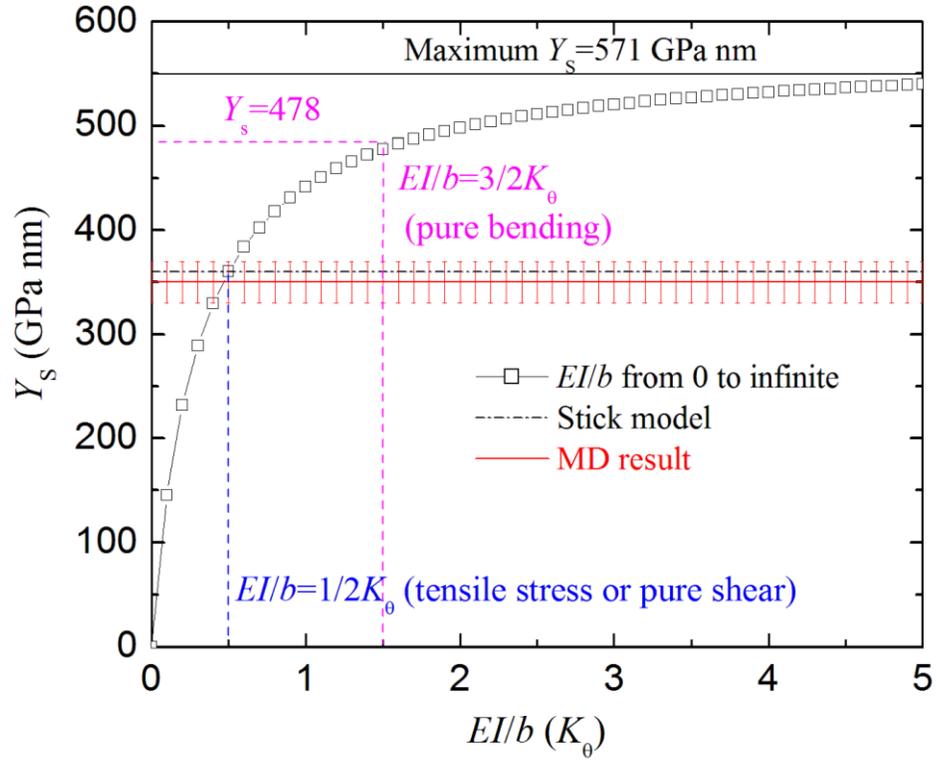

Fig. 4

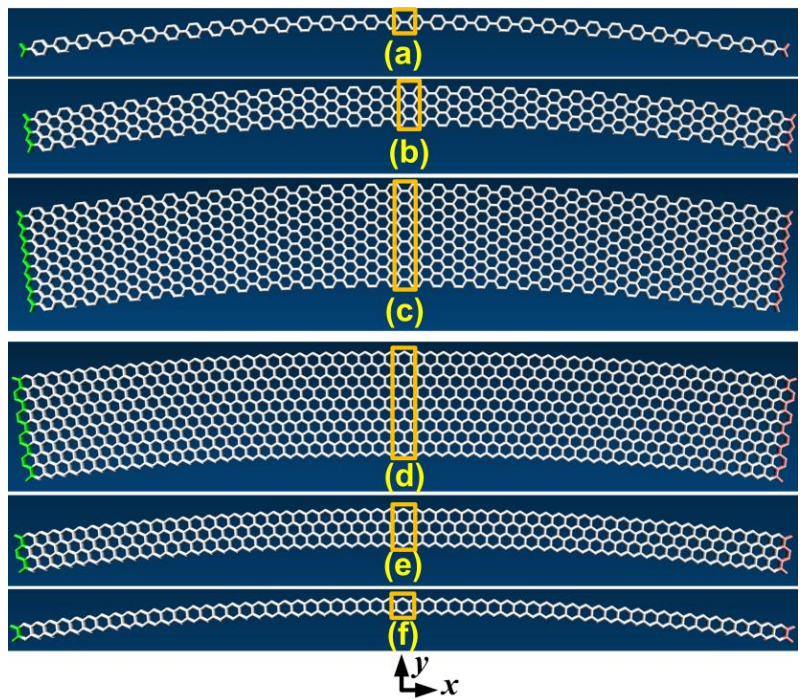

Fig. 5

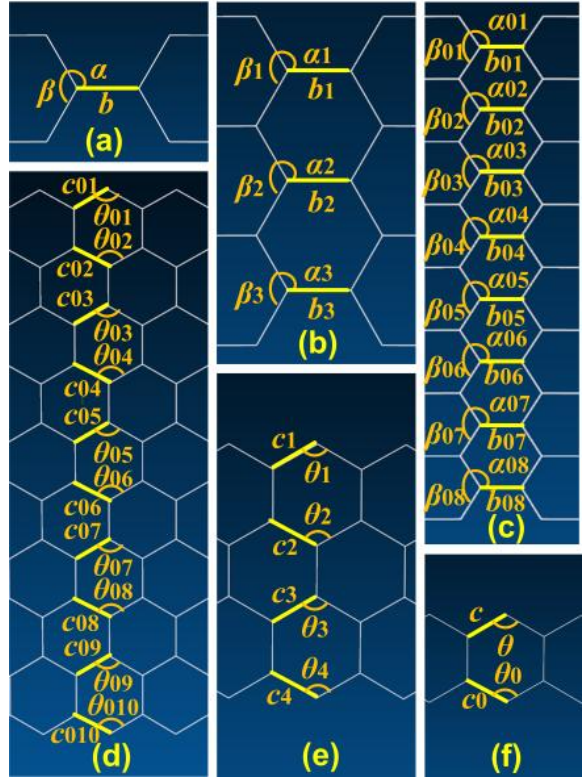

Fig. 6

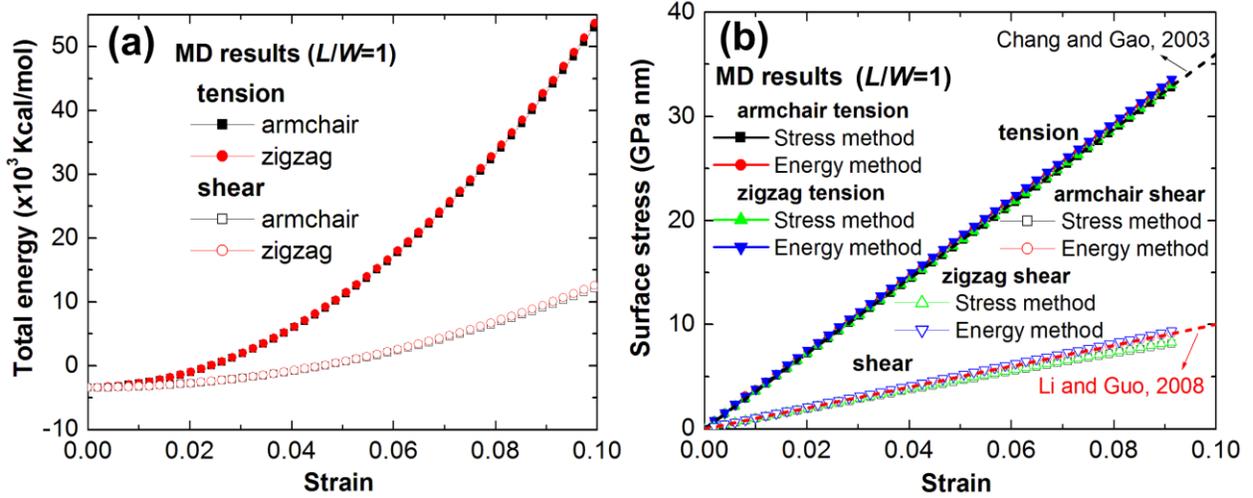

Fig. 7

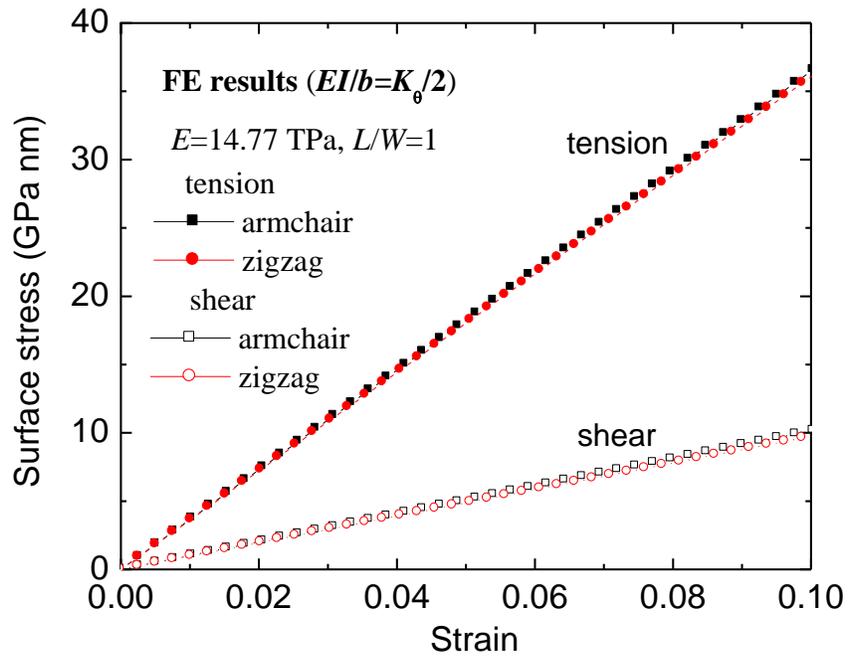

Fig. 8

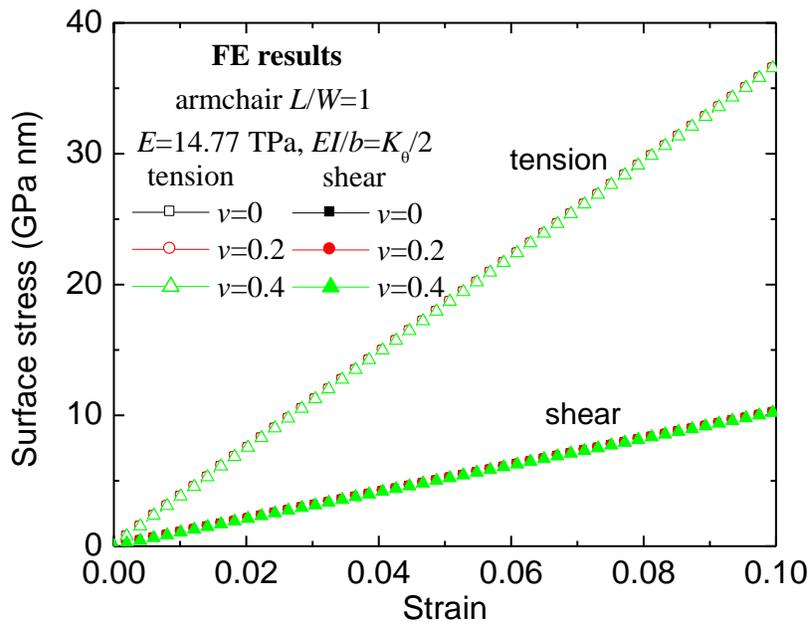

Fig. 9

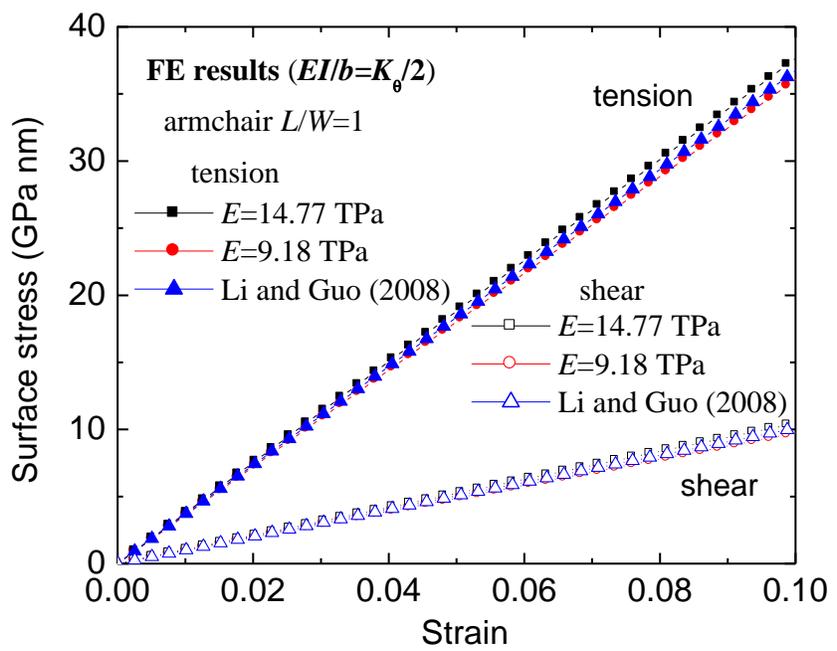

Fig. 10

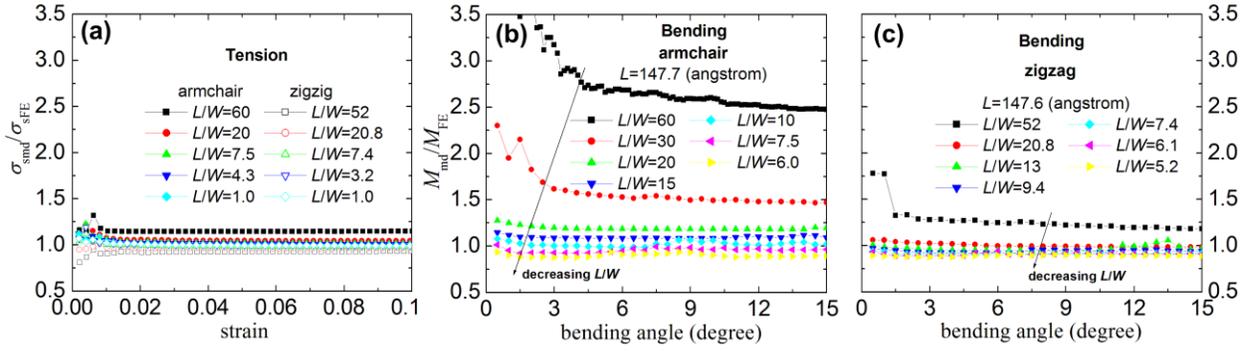

Fig. 11

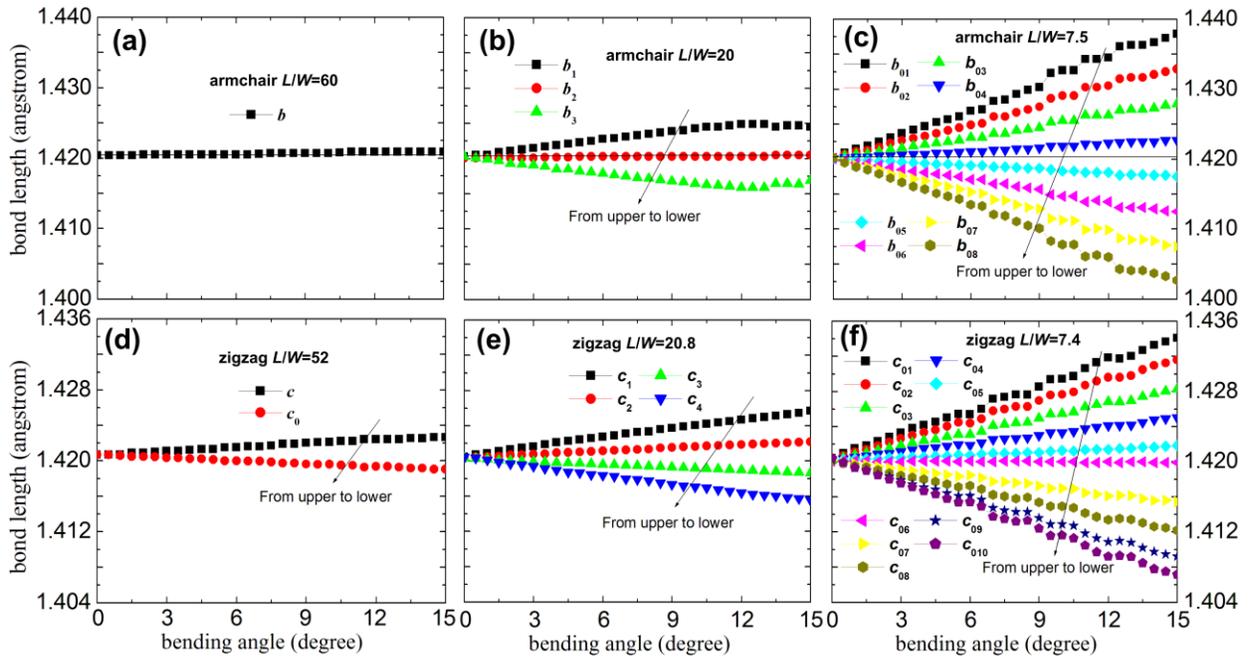

Fig. 12

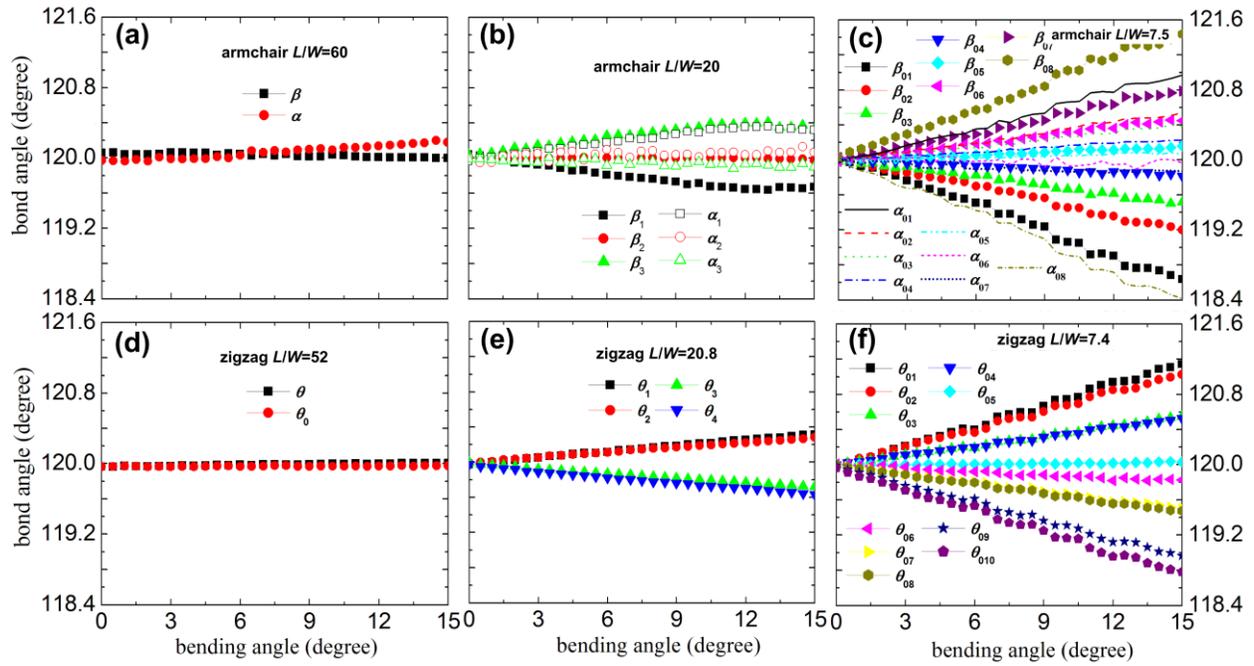

Fig. 13

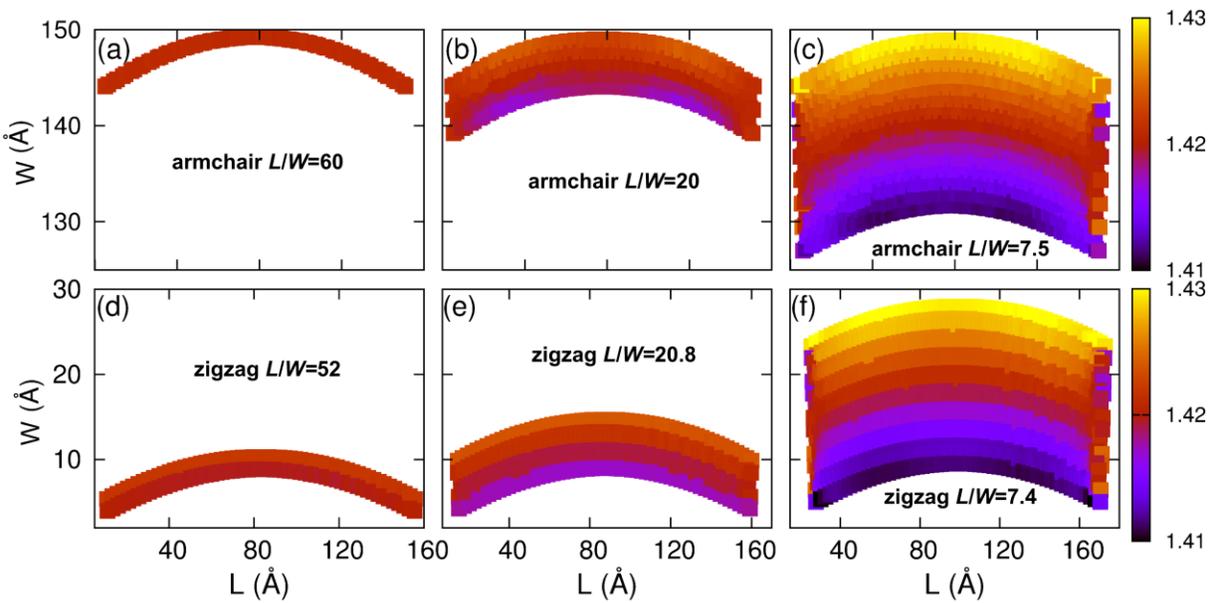

Fig. 14

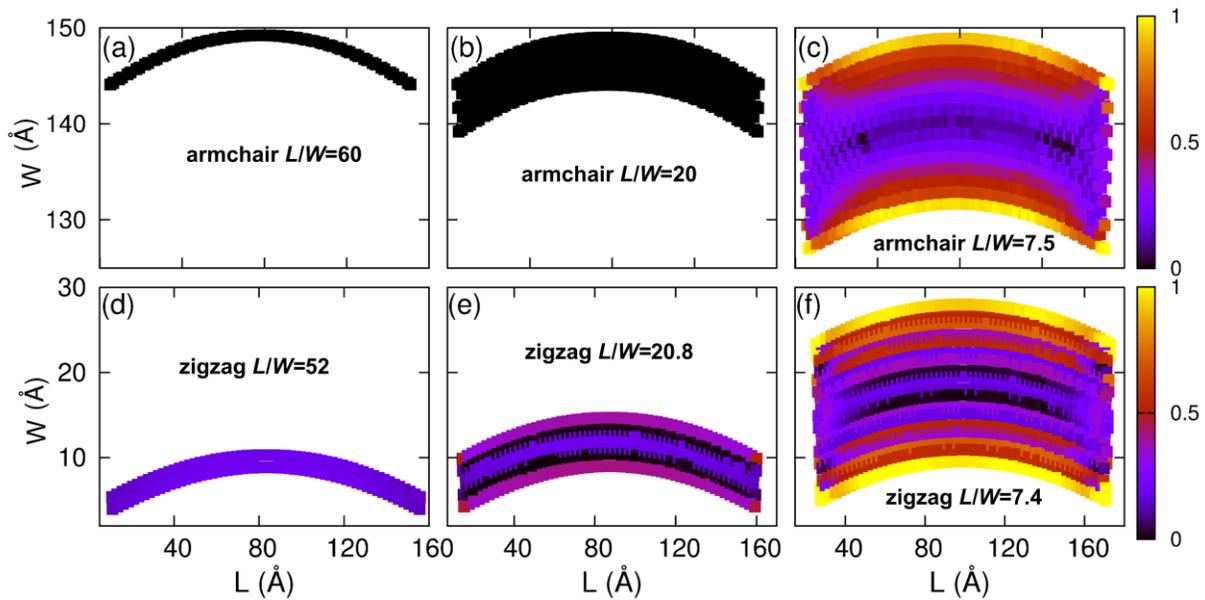

Fig. 15

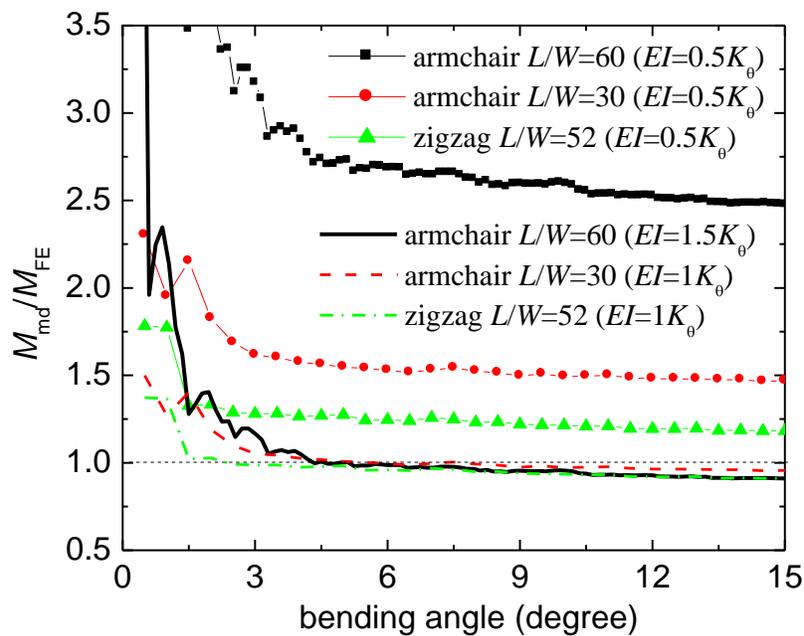

Fig. 16

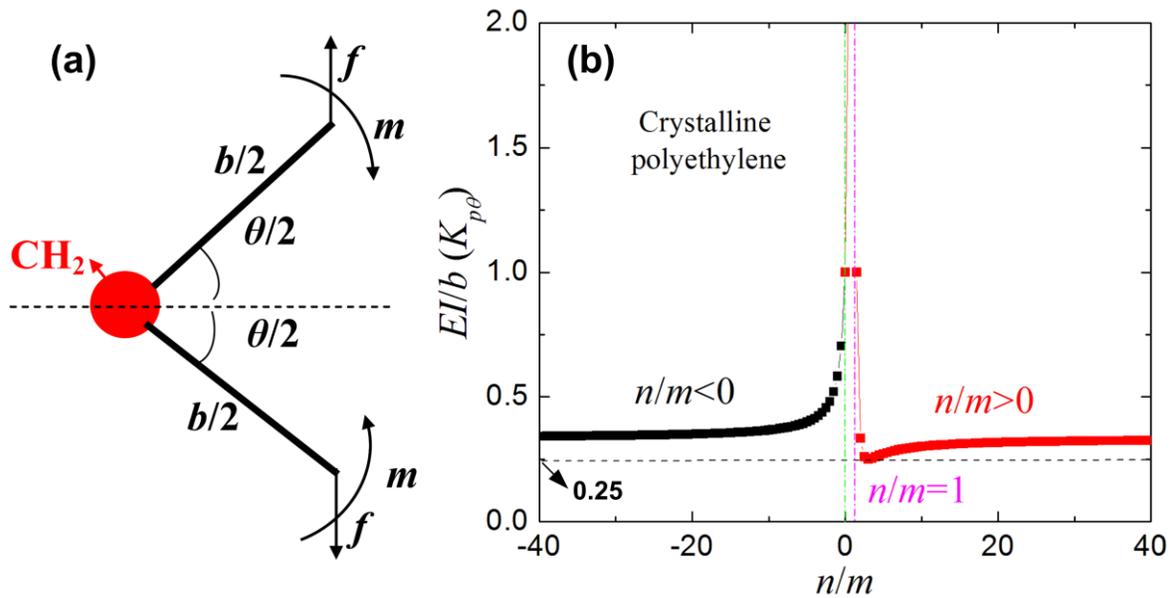

Fig. 17

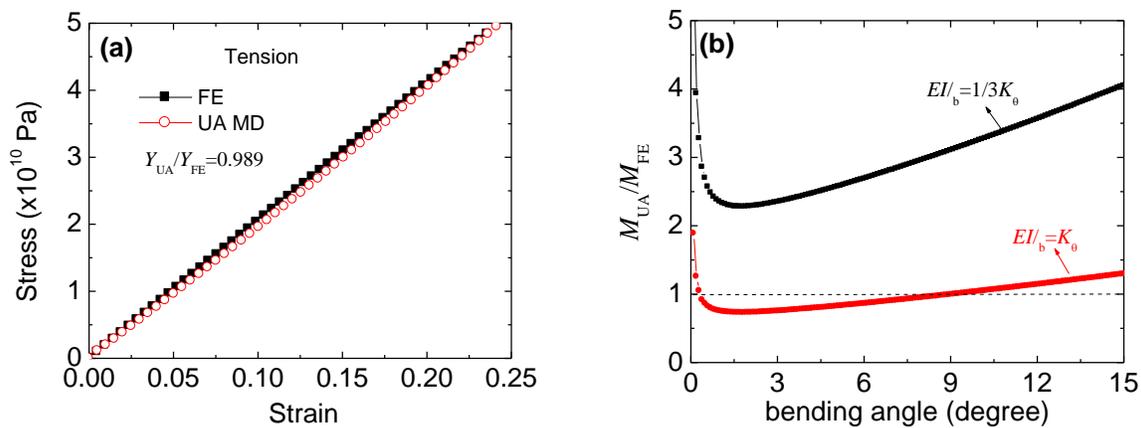

Fig. 18

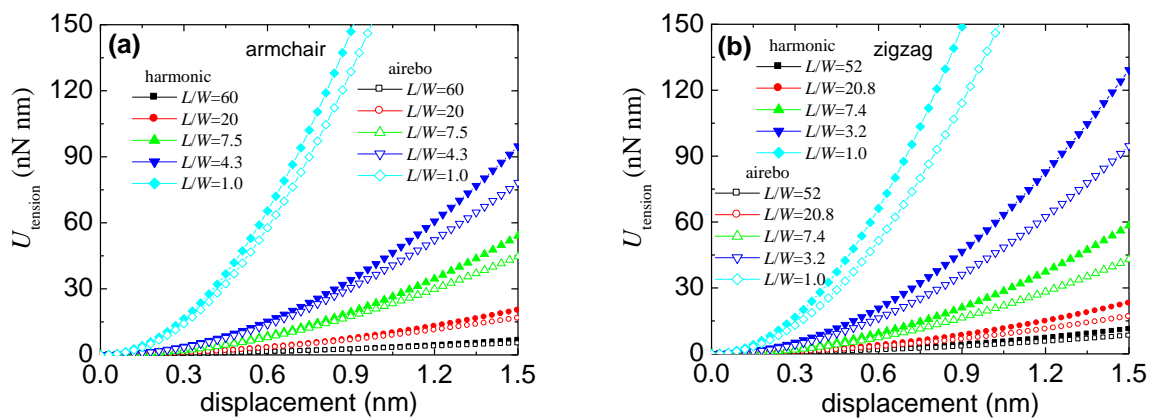

Fig. 19

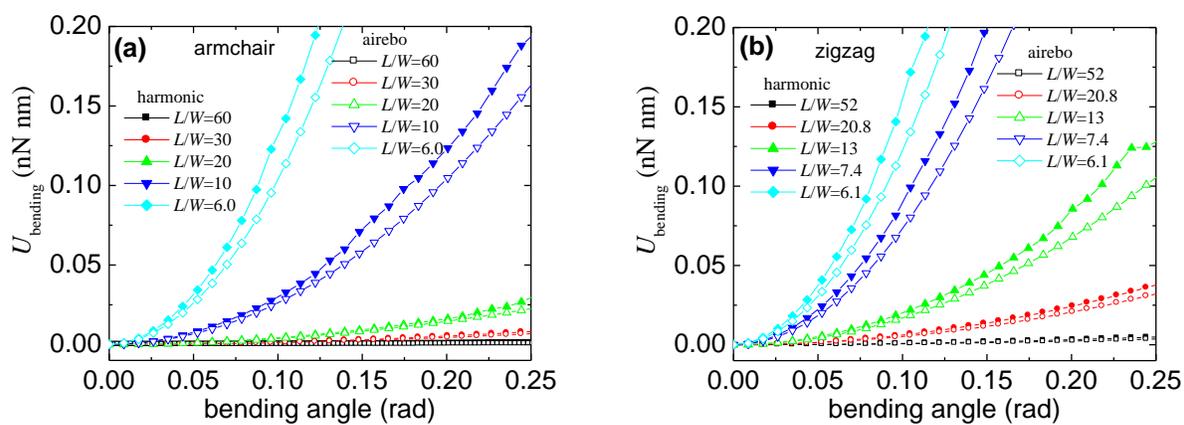

Fig. 20

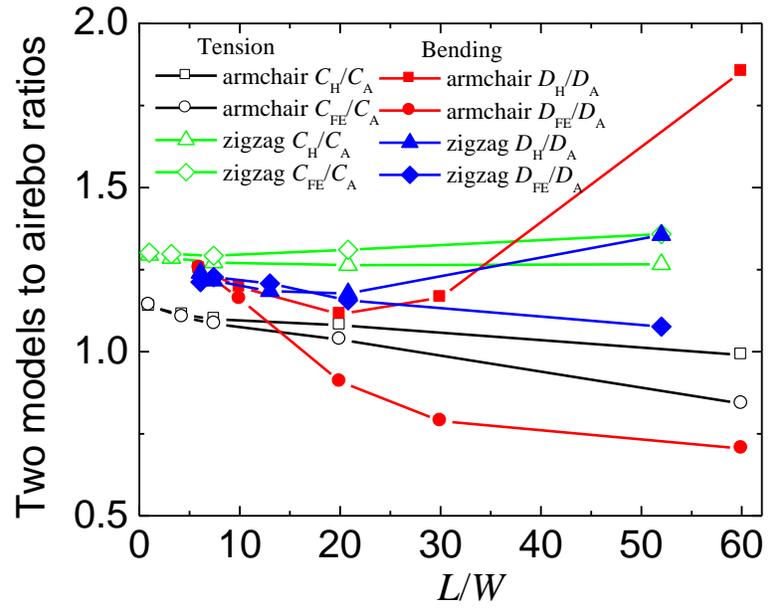

Fig. 21